\documentclass[namedreferences]{SolarPhysics}
\usepackage[optionalrh,solaenum]{spr-sola-addons} 
\usepackage{graphicx}                    
\usepackage{color}                       
\usepackage{url}                         

\begin{document}

\begin{article}

\begin{opening}

\title{The amplitude of sunspot minimum as a favorable precursor for the prediction of the amplitude of the next solar maximum and the limit of the Waldmeier effect}

%
\author{\surname{K. B. Ramesh}$^{1}$
        and \surname{N. Bhagya Lakshmi}$^{2}$      
       }

%
\runningtitle{Solar cycle prediction and the limit of the Waldmeier effect}

%
  \institute{$^{1}$ Indian Institute of Astrophysics, Koramangala, Bangalore-560034
                     email: \url{kbramesh@iiap.res.in} \\ 
             $^{2}$ Jyoti Nivas Pre-University \& Degree College, Koramangala, Bangalore - 560095
             }

\begin{abstract}
The linear relationship between the maximum amplitudes (R$_{max}$) of sunspot cycles and preceding minima (R$_{min}$) is one of the precursor methods used to predict the amplitude of the upcoming solar cycle. In the recent past this method has been subjected to severe criticism.  In this communication we show that this simple method is reliable and can profitably be used  for prediction purposes.  With the 13-month smoothed R$_{min}$ of 1.8 at the beginning, it is predicted that the R$_{max}$ of the ongoing cycle will be around 85$\pm$17,  suggesting that Cycle 24 may be of moderate strength.   Based on a second order polynomial dependence between the rise time (T$_R$) and R$_{max}$, it is predicted that Cycle 24 will reach its smoothed maximum amplitude during the third quarter of the year 2013.  An important finding of this paper is that the rise time cycle amplitude relation reaches a minimum at about 3 to 3.5 years corresponding to a cycle amplitude of about 160. The Waldmeier effect breaks at this point and T$_{R}$ increases further with increase in R$_{max}$. This feature, we believe,  may put a constraint on the flux transport dynamo models and lead to more accurate physical principles based predictions.
\end{abstract}

%

\end{opening}

%
 \section{Introduction}
Various methods have been used in the past to predict the amplitude (R$_{max}$) and time of maximum of a  sunspot cycle.  Some of them are based on sound physical principles \cite{Dikpati06,Choudhuri07}.  While a wide range of predictions have been made (See reviews by \opencite{Pesnell08}, \opencite{Brajsa09}, \opencite{Hathaway10}, \opencite{Ahluwalia10}, \opencite{Petrovay10}, \opencite{Kakad11}) for the upcoming Solar Cycle 24, very few of them use solar precursors \cite{Schatten05,Svalgaard05,Javaraiah07}.  The importance of using solar precursors is that their physical explanations could have implications for dynamo models. One of the useful prediction methods, using solar precursors,  is the dependence of R$_{max}$ on the preceding minimum, R$_{min}$ \cite{Brown76,Hathaway02,Hathaway10}. The relation between the rise time (the time between the occurrence of minimum and the following maximum, T$_R$) and  R$_{max}$  \cite{Waldmeier35} was used to predict the time of R$_{max}$ of the upcoming cycle.  Earlier these methods have  been applied to predict R$_{max}$ and T$_R$ of Solar Cycle 23 \cite{Ramesh00}.  If not a complete contradiction, limitations of these methods were expressed with particular reference to individual cycles \cite{Wang09}. 

Use of R$_{max}$ versus R$_{min}$ (low correlation), among other precursor methods (see \opencite{Petrovay10} for a detailed review on solar cycle prediction), for prediction purposes has taken a back seat because of high correlation values of R$_{max}$ with other precursors such as geomagnetic index, aa$_{min}$ (\opencite{Wilson98}) for single variate and combination of aa$_{min}$ and R$_{min}$ \cite{Kane97,Wilson98} for bivariate models. However,  in the recent solar cycle  these models could not perform well with their predictions (See \opencite{Dl09};  \opencite{Du11} for explanation) while some of them could do reasonably well \cite{Ahluwalia00}.   It is therefore very important to look for a precursor that has real physical significance to the amplitude of the next cycle.  The strength of the polar field during solar minimum is one of such useful precursors \cite{Schatten78,Choudhuri07}.  The polar fields reach their maximal amplitude near minima of the sunspot cycle, and hence the prediction becoming available 2-3 years before the upcoming maximum.  From the known fact that the poloidal field generate the toroidal field of the next solar cycle, and that the parameter, R$_{min}$, being a proxy for the toroidal fields, can serve as a precursor. The advantage of using the strength of the polar field and R$_{min}$ is that these quantities are directly linked to the amplitude of the next cycle through the Sun's internal dynamics while the geomagnetic and interplanetary precursors are the quantities representing the effects of solar activity.   R$_{min}$, being more easily observable quantity compared to the  strength of the polar field, may be a preferable one. This method, however, requires that the minimum epoch is already known, and that the method can be applied only some time into the new cycle, when proper averaging can be done to define the minimum in view of strong fluctuations of activity around minimum \cite{Harvey99}.

Owing to the overlapping of sunspot cycles, quantities related to solar minima as precursors for predicting the amplitude of the next solar cycle needs to be dealt with care \cite{Cameron08}.  In a recent study, \inlinecite{Du10} opined that the relationship between R$_{max}$ and R$_{min}$ is insufficient to infer the amplitude of the upcoming cycle. In this context, we present the results of our analysis that supports the usefulness of this method in predicting R$_{max}$ of the upcoming cycle.

\section{Data}

In this study, we use sunspot number data (Rz) obtained from ftp://ftp.ngdc. noaa.gov/STP/SOLAR\_DATA/SUNSPOT\_NUMBERS/INTERNATIONAL/\newline monthly/MONTHLY.PLT for the duration January 1749 to November 2010 covering 23 solar cycles. R$_{max}$ and R$_{min}$ of the solar cycles are deduced from 13-month running means of 3-month smoothed monthly sunspot number.  

%
 \begin{figure} 
 \centerline{\includegraphics[width=1.0\textwidth,clip=]{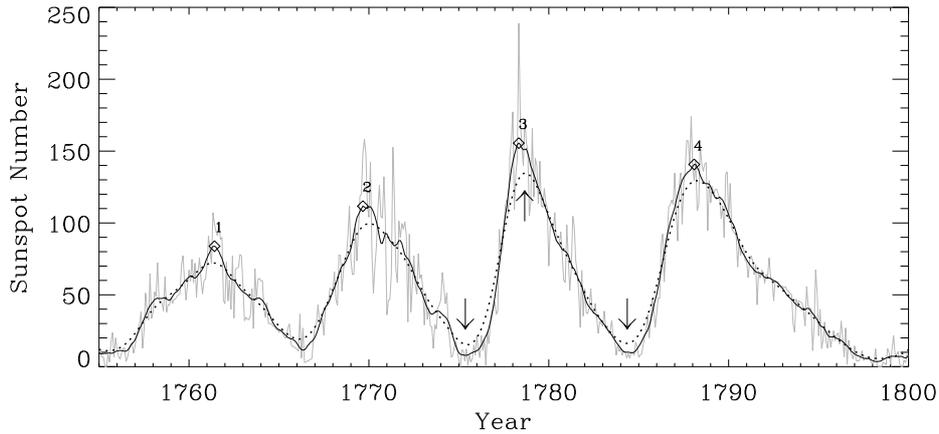}}
 \caption{Thin line depicts the  Monthly averaged sunspot number for Cycles 1 to 4. Thick line represents the 13-month running averages obtained from 3-month smoothed monthly averages.  Dotted line shows the five times successively smoothed monthly averages with 13-month window. Cycle numbers are shown at the peak of every cycle.   Few examples of large deviations in R$_{max}$ and R$_{min}$ due to excessive smoothing (dotted line) are marked with arrows.}
 \end{figure}

\section{Maxima and minima of solar cycles}
Solar activity is inherently noisy and hence determining maxima and minima and dates of their occurrences is a difficult task. Monthly averaged daily sunspot number is commonly used to study the long term behavior of solar activity and for prediction purposes. Traditionally 13-month running window was used to smooth the data prior to determining R$_{max}$ and R$_{min}$ while other methods of smoothing also have been attempted earlier (Hathaway, 2010 and references therein). We use this method to determine R$_{max}$ and R$_{min}$ but with an additional smoothing with 3-month running window prior to applying 13-month smoothing.  We observed that a preliminary smoothing with 3-month window avoids ambiguity in determining R$_{max}$ and R$_{min}$ among the multiple peaks, particularly seen during times of maximum and minimum epochs within a solar cycle. We noticed that any further preliminary smoothing show appreciable changes not only in R$_{max}$ and R$_{min}$ but also in their timings. Figure 1 shows a typical example of smoothing (only Cycles 1 to 4 are shown for clarity) wherein the smoothed version of the profile (thick continuous line) show maxima and minima at and around the average of times of extremes (Mckinnon, 1987) reached in Rz  (shown with a faint continuous line) while excessive smoothing (dotted line) show large deviations (indicated with arrows at the beginning, maximum and end of the cycle 3) in both R$_{max}$ and R$_{min}$ and their respective time of occurrences.  This method of optimizing the smoothing effect for obtaining unambiguously the R$_{max}$ and R$_{min}$ values, we believe, is not unique and other methods may work as well.

%
 \begin{figure} 
 \centerline{\includegraphics[width=0.6\textwidth,clip=]{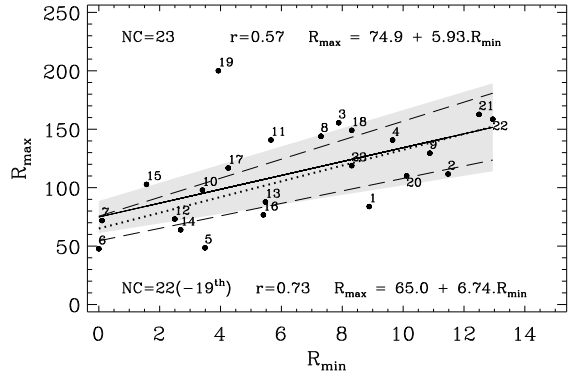}}
 \caption{Scatter plot of R$_{max}$ against R$_{min}$. Thick continuous line shows the linear fit for all the 23 cycles included.  Hatched region show 1 $\sigma$ level uncertainty.  Dotted line represents the regression line for 22 cycles with R$_{max}$ of cycle 19 excluded.  Region embedded between two dashed lines show the respective 1$\sigma$ level uncertainty. Respective correlation coefficients and regression equations are also shown in the figure. Note the near overlap of the two uncertainty regions.}
 \end{figure}

\subsection{R$_{max}$ versus R$_{min}$}

Figure 2 depicts the scatter plot of R$_{max}$ Versus R$_{min}$.  The correlation (correlation coefficient r = 0.57)  is significant at a confidence level (CL) of 99\%. Probable error of correlation (PE = 0.6745 $\times$ [1-r$^2$]/$\sqrt{nc}$, where nc is the number of cycles) is 0.095.  The correlation coefficient being greater than 6 times PE, R$_{max}$ is supposed to be related to R$_{min}$ with a high degree of correlation. The test statistic, t (r $\times$ $\sqrt{(n-2)/(1-r^2)}$) = 3.187 (t value for 99\% CL is 2.831) also indicates the correlation to be significant at 99\% CL.  However, the coefficient of determination (r$^2$ = 0.325) indicates that nearly two thirds of the variance in R$_{max}$ is unexplained by the correlation.  It is to be noted that a relationship can be strong and yet not significant numerically. Conversely, a relationship can be weak but significant. Therefore, we further carry out the regression analysis to check for real statistical significance between them that also help predicting the amplitude of the upcoming solar cycle.

Continuous line in Figure 2 depicts the linear fit of the form R$_{max}$ = A + B.R$_{min}$, ( A - intercept on the ordinate and B - slope of the fitted line). The test statistic t$_s$ $ [(B-0.0)/\sqrt{(MSE/S_{xx}})$ where MSE is the mean square error in R$_{max}$ and S$_{xx}$ is the mean square error in R$_{min}$] for the regression coefficient B turns out to be 3.2. Similar regression between R$_{max}$ and R$_{min}$ with Cycle 19 excluded is shown with dotted line and the corresponding t$_s$ is 4.8.  In both these cases the test of significance (99\% CL) indicate strong dependence of R$_{max}$ on R$_{min}$. 

\subsection{Trends in recent cycles}

Recent studies show that the prediction relies more on the recent cycle than on the far past ones \cite{Schatten05,Svalgaard05,Du11}. \inlinecite{Du10} have claimed that the correlation between R$_{max}$ and R$_{min}$ is mostly contributed by the cycles 1-14 and argued that the cycles 15-19 behave differently from the earlier cycles.
Figure 3a shows the scatter plot of R$_{max}$ against R$_{min}$ for the cycles 1 to 14. The regression line (R$_{max}$ = 60.6 + 6.92 . R$_{min}$) seems to be comparable to the line shown in Figure 2 for all the 23 cycles.  Correlation coefficient (0.71, CL=99\%) is higher than the overall correlation coefficient of 0.57.  In contrast, for the cycles 15-23 (Fig 3b) the correlation coefficient has decreased to 0.28 (not significant even at 90\% CL) and the regression line is completely different (R$_{max}$ =112.7 + 2.70 . R$_{min}$).  However, it is interesting to note that a regression line (R$_{max}$ = 81.2 + 5.45 . R$_{min}$) for the cycles 15-23 excluding the Cycle 19 is a close match to that of Cycles 1-23 ( R$_{max}$ = 74.9  +  5.93 . R$_{min}$ ) and the correlation (0.73, CL=95\%) is almost similar to those of Cycles 1-14 (0.71) and Cycles 1-23 (0.73).  Hence it is clear that but for the Cycle 19 the trend in the behavior of amplitudes of recent solar cycles is similar to that of the earlier cycles. 

%
 \begin{figure} 
 \centerline{\includegraphics[width=1.0\textwidth,clip=]{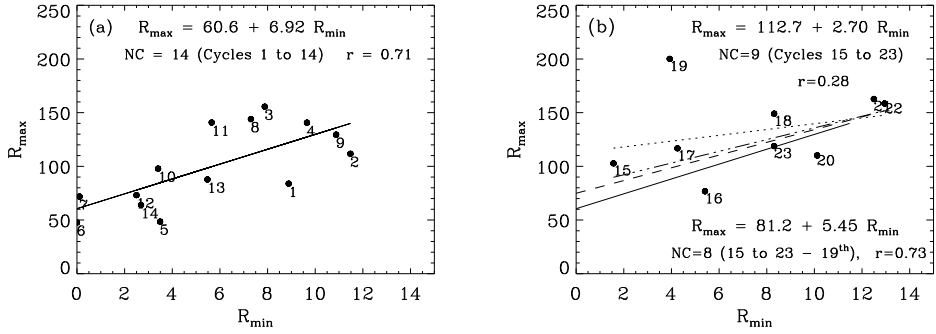}}
 \caption{(a) Scatter plot of R$_{max}$ against R$_{min}$ for cycles 1 to 14 and the corresponding regression line.  Panel (b) show similar diagram for the Cycles 15 to 23 and the corresponding linear fit (dotted line). Dash-dot-dot-dot line depicts the linear fit of Cycles 15 to 23 without Cycle 19. The corresponding regression equation and the correlation coefficient are shown at the lower portion of panel (b). The thick continuous line and the dashed line represent respectively the regression lines for Cycles 1 to 14 (as shown in panel (a)) and for Cycles 1 to 23 (as shown in Fig 2) for comparison.}
 \end{figure}

In order to confirm this result we have carried out further analysis using progressive correlations and regressions.  Top panel of Figure 4 shows the progressive correlation coefficients (continuous line) of R$_{max}$ and R$_{min}$ for the Cycles 1-6, 1-7, ..... , 1-23.  Similar curve (dotted line) excluding Cycle 19 is also shown in the figure. The dotted line is plotted with a small vertical offset in order to identify the curves clearly at the locations of overlapping. The variation in the correlation coefficient seems to be consistent up to Cycle 18 and  varied between 0.65 and 0.75.  A drop in the correlation coefficient (indicated  by arrows in top panel of Fig 4) for Cycles 1-19 and beyond is quite apparent. However,  they are significant to the level of 99\%.   When Cycle 19 is excluded from the analysis, similar trend (dotted line) as seen in cycles up to 18 continuous for Cycles 1-20 and beyond.  In fact the correlation remains all through 23 cycles between 0.7 and 0.8 when Cycle 19 is not included in the analysis.   Similar trends are apparent in both the regression coefficients (A and B in the top panel of Fig 4) when cycle 19 is not included.  Therefore, this trend of Cycles 15-23, except the Cycle 19, rules out the view that they behave differently when compared to the behaviour of Cycles 1-14.

We further demonstrate the effect of Cycle 19 on the correlations of all the 23 cycles and the recent nine cycles using temporal variation in the running correlations \cite{Du10}. Curve (continuous line) depicting the running correlation,  r(5,n), of R$_{max}$ and R$_{min}$ evaluated with a moving time window of 5 cycles is shown in the bottom panel of Figure 4. It is quite apparent that the correlations become negative beyond Cycle 18 \cite{Du10}.  However the correlations beyond Cycle 18 also become positive (dotted line between Cycles 16 and 21 shown in the bottom panel) when Cycle 19 is eliminated from the analysis. Therefore it is clear that the negative correlations seen beyond Cycles 18  arise only due to the anomolous behaviour of Cycle 19 and that the recent cycles (except Cycle 19) behave in a way more or less similar to earlier ones (1-14 cycles).

%
 \begin{figure} 
 \centerline{\includegraphics[width=0.6\textwidth,clip=]{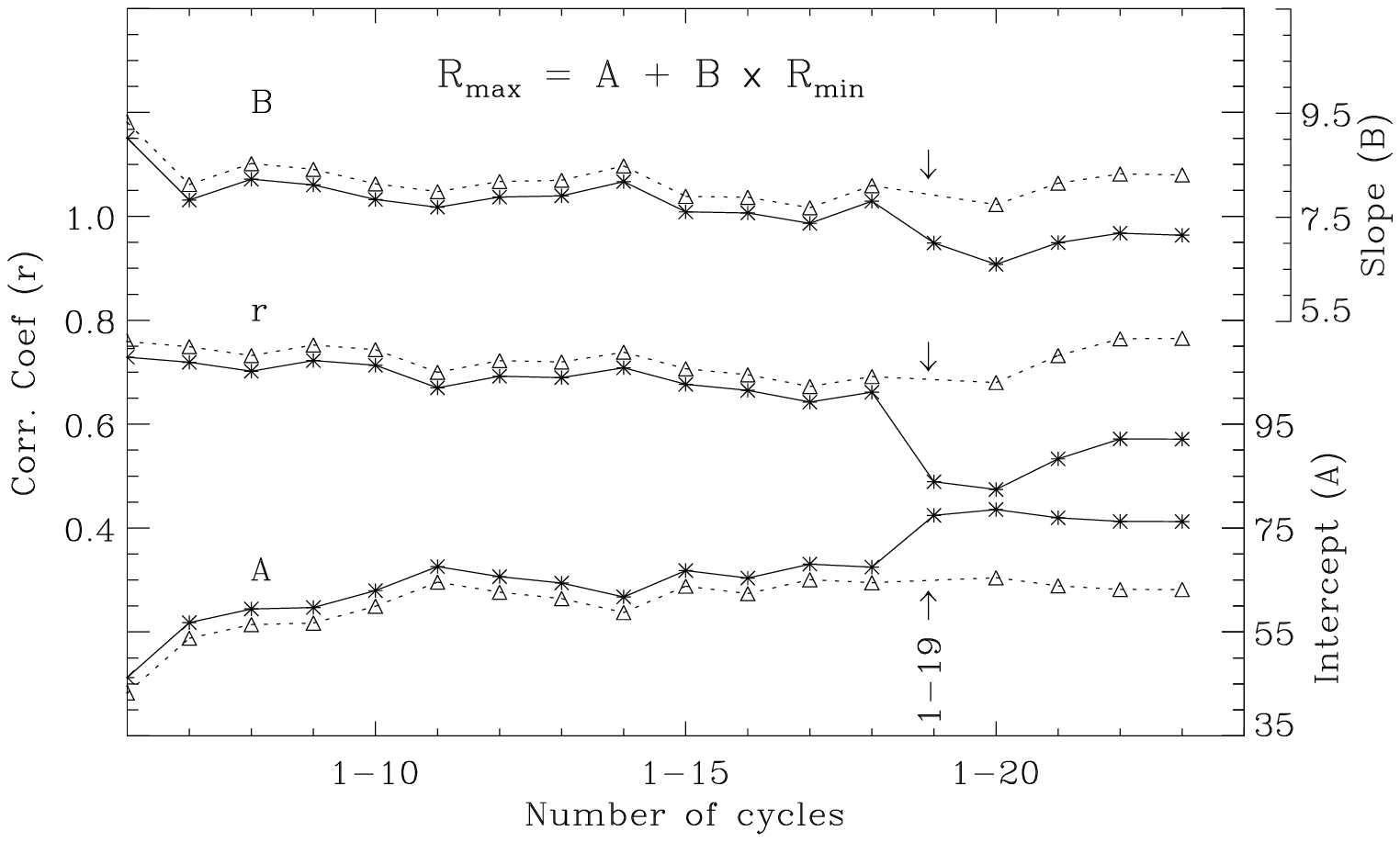}}
\vskip .3cm
\hskip 2.2cm  {\includegraphics[width=0.555\textwidth,clip=]{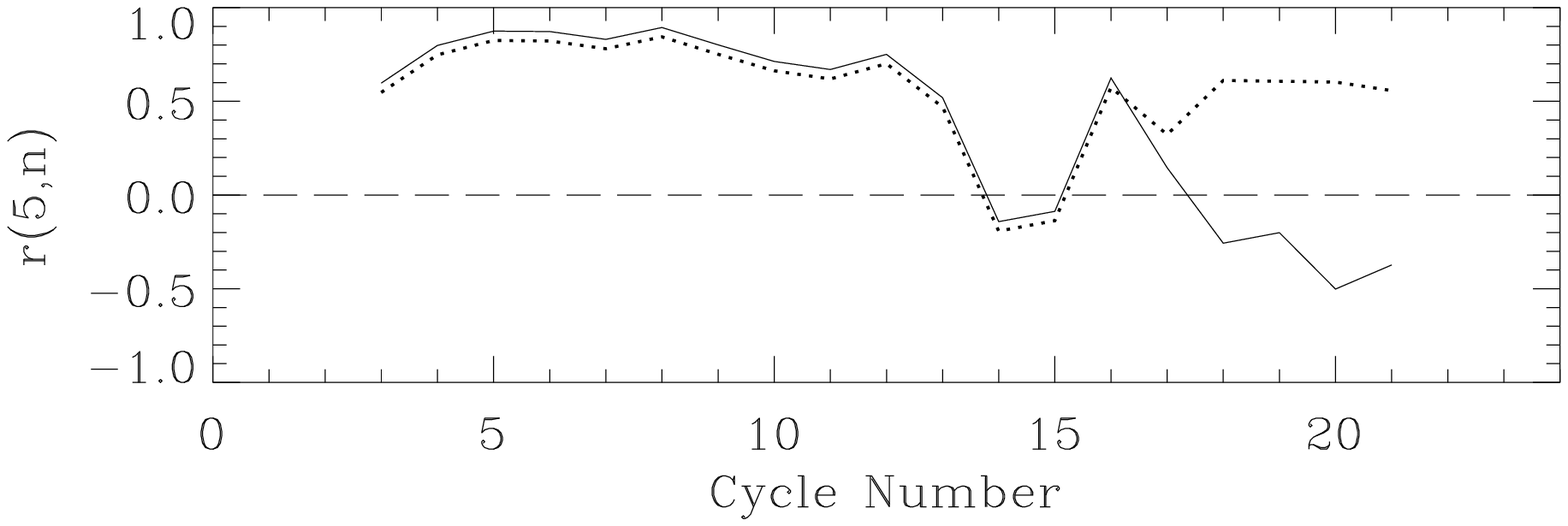}}

 \caption{Top panel: Continuous lines are the plots of progressive correlation (r) and regression (A and B) coefficients of R$_{max}$ versus R$_{min}$. Dotted lines show similar plots with cycle 19 excluded in the analysis. Dotted lines are plotted with a small vertical offset to identify the lines clearly at the locations of overlap. Bottom panel: Continuous line shows the running correlations r(5,n) of R$_{max}$ versus R$_{min}$ with 5 cycle window. Dotted line shows similar curve with cycle 19 excluded from the analysis. Note the switching of r(5,n) from negative (continuous line) to positive (dotted line) at cycles 17, 18, 20, 21. }
 \end{figure}

\subsection{Cycle 19 - really an outlier?}

Above analysis show that the regression line of R$_{max}$ versus R$_{min}$ for Cycles 1-23 without Cycle 19 does not vary significantly from that of the Cycles 1-23 with all cycles included (Fig 2).   Fitted value of R$_{max}$ for Cycle 19 with R$_{min}$ of 3.7 is 99 while its observed R$_{max}$ value is 202. However, for similar R$_{min}$ values (e.g.,  3.3 and 3.8 respectively for cycles 10 and 17) the fitted values (96.5 and 99.6) seem to be very closely matching with those of observed ones (98 and 118).  Thus the observations such as that of Cycle 19 are often called as 'outliers' (Kane, 2007).

We check this behavior of Cycle 19 using Chauvenet's criterion \cite{Kennedy64}  for naming an observation as an outlier.    According to this criterion, an observation is rejected from a sample if the coefficient of outlier (Oc = [$x_i$-$\overline{x}$]/$\sigma$, where $x_i$ is the observation under consideration, $\overline{x}$ -  mean of the sample and $\sigma$ -  the standard deviation of the sample) is greater than the tabulated value for a particular sample size.  With the R$_{max}$ (202.6) for Cycle 19, mean of R$_{max}$ (113.9) for all the 23 cycles and the corresponding standard deviation (40.25), Oc turns out to be 2.204.  This value is smaller than the tabulated (Table A-6, \opencite{Kennedy64}) value (2.33) for 23 samples  and hence cannot be accounted for the rejection of Cycle 19 from the sample. We further test this hypothesis by considering the recent Cycles 15-23.  With the R$_{max}$ (202.6) for Cycle 19, mean of R$_{max}$ (134) for the Cycles 15-23  and corresponding standard deviation (38.0), Oc turns out to be 1.805. Table value for 9 observations being 1.91, does not allow the Cycle 19 to be knocked out of the sample of 9 cycles.   Therefore, naming the solar Cycle 19 as an anomalous may be more appropriate than calling it as an outlier and deserves further attention. "It is to be noted that the decision to reject an observation be made on the basis of experience and must not be made lightly.  It is important to realize that in rejecting an observation, we may be in our ignorance, throwing away vital information which could lead to the discovery of a hitherto unrecognized factor"  \cite{Kennedy64}.  

\subsection{Trends of variation in consecutive cycles}

Through an analysis of trends, V$_m$ [sign\{R$_{max}$(i) - R$_{max}$(i-1)\}] and V$_{min}$ [sign \{R$_{min}$(i) - R$_{min}$(i-1)\}] \inlinecite{Du10} have supported the view that the  R$_{max}$ versus R$_{min}$ relation is not always effective for individual cycles \cite{Wang09} and that the negative correlation between V$_m$ and V$_{min}$ cannot account for the predicted (\opencite{Pesnell08} and references therein) very weak R$_{max}$ for the very low R$_{min}$.  Based on the analysis of temporal variation in the running correlations they have indicated that a  lower R$_{min}$ has not always been followed by a weaker R$_{max}$ and therefore inferred that Cycle 24 need not be a very weak cycle.

It is pertinent to mention here that the trends in variations  in both the parameters (R$_{max}$ and R$_{min}$) under consideration for the correlation analysis need not be the same. Having V$_m$ and V$_{min}$ \cite{Du10} same sign implies a systematic error involved in their measurements. In particular,  if V$_m$(i) proportional to V$_{min}$(i),  then  R$_{max}$ versus R$_{min}$ will be  monotonous (increase in case of positive V$_m$ and V$_{min}$ or decrease in case of negative V$_m$).   It is to be noted that not only instrumental errors but also the atmospheric “seeing” errors play a crucial role in determining  R$_{max}$ and R$_{min}$.  More over the dynamics involved in the evolution of the solar activity from quiet to maximum situation in a magnetically complex system is still a little understood process. Therefore, random errors leading to great scatter in the diagram, are expected in such observational data.  We, therefore opine that building good statistics with more number of data points may lead to tight correlation between them.   However, it is not too small a data set to work on as of now.  It is to be noted that the statistical trend in R$_{max}$ does not vary drastically when the anomalous Cycle 19 (not an outlier) is included in the analysis.  The regression (continuous line in Figure 2) coefficient is 5.93$\pm$1.85  for all the 23 cycles while it (dashed line in Figure 2) is 6.74$\pm$1.40 when Cycle 19 is excluded from the analysis. This feature can also be seen from the uncertainties in the regression lines of 1-23 cycles with Cycle 19 included (hatched region centered at the regression line shown with continuous line) nearly coinciding with that of the Cycles 1-23 without Cycle 19 (region embedded between two dashed lines centered at the regression line shown in dotted line) in Figure 2.  With the support of the tests performed,  we opine that the currently available 23 cycles data show statistically significant relationship between R$_{max}$ and R$_{min}$ and that this relationship, within the mentioned uncertainty levels,  can be used for predictions.  

\subsection{Prediction of R$_{max}$}

The regression line (R$_{max}$ = 76.26 + 6.18 R$_{min}$) obtained using R$_{max}$ {\it versus} R$_{min}$ (r=0.57) relation for 22 cycles has been used to predict a maximum value of 126 $\pm$ 26 for the solar cycle 23 \cite{Ramesh00}  that was very close to the observed value of 120.8.  The linear regression equation for the observed data of all the 23 cycles (continuous line in Fig 2) is R$_{max}$ = 74.9($\pm$13.7)  +  5.93($\pm$1.85).R$_{min}$.   This regression equation, with R$_{min}$ of 1.8 (13-month running average of 3-month smoothed monthly sunspot number) at the beginning of the upcoming Cycle 24, gives an estimation of R$_{max}$ to be 85$\pm$17. The uncertainties provided in the estimation is derived based on the uncertainties in the regression coefficients.

When the progressive correlations are considered the correlation between R$_{max}$ and  R$_{min}$ (top panel of Figure 4) is quite consistent even though the overall correlation value remains low. It is to be noted that the effect of anomalous behaviour of Cycle 19 and the short-term negative correlations [as seen in r(5,n)] on the relationship of R$_{max}$ and  R$_{min}$ reduce with increased number of data points in the sample.  Earlier studies (eg., \opencite{Cameron08}, \opencite{Du11} and references therein) indicated that the high correlation values need not always yield an accurate prediction. Consistency in the correlation between R$_{max}$ and  R$_{min}$ seems to be an important factor and the slow recovery (Figure 4 top panel) towards the consistent value after the anomalous Cycle 19 seems to provide a greater strength to the present linear trend (Figure 2) seen between them.

\section{Cycle Rise Time}

From Figure 1 and from the profile of all the 23 solar cycles (not shown) it is quite clear that the time of rise of the cycle is shorter than that of the decay. Solar activity maxima occur 3 to 4 years after the minimum, while it takes about 7 to 8 years to reach next minimum. The length of the rise phase appear to decrease with increase in cycle maximum, while the length of the decay phase does not show such relationship.

\subsection{T$_R$ versus R$_{max}$}

Predicting the time of maximum of the upcoming solar cycle is an issue discussed for several decades. It was first \inlinecite{Waldmeier35} who formulated the inverse correlation between rise time and the cycle amplitude. \inlinecite{Lantos00} and \inlinecite{Cameron08} have pointed out that the high correlation between the rate of rise of the cycle and R$_{max}$ can serve as a better tool to predict the time of maximum. \inlinecite{Du09} found that the Waldmeier effect is very weak for some periods of time and the long term varying behaviour of the correlations represents an observational constraint on solar dynamo models. \inlinecite{Lantos00} opined that the anticorrelation between T$_{R}$ and R$_{max}$ cannot be used to predict the time of maximum because the time of maximum is already known by the time maximum occurs.  However,  R$_{max}$ prediction obtained from the linear relationship of R$_{max}$ and R$_{min}$ is more often used in the relationship of T$_{R}$ and R$_{max}$ to predict the T$_R$ of an upcoming cycle. Other methods such as inverse correlation between R$_{max}$ and the length of the cycle \cite{Hathaway94},  R$_{max}$ and the length of the third preceding cycle \cite{Solanki02}  also have been introduced. In most of the rise-time modeling an inverse linear regression fit has been used to establish a relationship between T$_R$ and R$_{max}$.  

In the mean while \inlinecite{Dikpati08} have made an attempt to explain that the Waldmeier effect is specific to only sunspot number and that this effect does not exist in the sunspot area data.  \inlinecite{Karak10} have opined that the analysis of \inlinecite{Dikpati08} has ended up with that result because of improperly defined rise time of a sunspot cycle. \inlinecite{Karak10}, by defining the rise time (T$_{RKC}$) as the time taken for the activity to grow from 0.2R$_{max}$ to 0.8R$_{max}$ have further explained that this effect truly exists even in sunspot area data and that the data needs to be handled carefully. However, it is out of the scope of this paper to discuss this controversy but to explain that this phenomenon can profitably be used for the prediction of the time of maximum if the data is handled in a more meaningful way.

Figure 5 shows the scatter plot of T$_R$ (in years) versus R$_{max}$.  A linear fit (Equation III in Figure 5) with correlation coefficient of -0.75 is shown with a dotted line.   A straight line (dashed line and equation IV) with a correlation coefficient of -0.81 for 22 cycles (excluding Cycle 19) is also shown in the Figure. Two regression lines do not differ much indicating that eliminating the anomalous Cycle 19 do not make much difference on the inverse relationship of T$_R$ versus R$_{max}$. However, visual inspection of the scatter plot clearly indicate that Cycle 19 is undoubtedly anomalous in the sense that the deviation of observed T$_R$ of Cycle 19 from those of the fitted lines is too high.    The linear regression equations  further indicate that with R$_{max}$ of about 202 the time of rise of Cycle 19 should have been as short as 2.5 years while the observed rise time (from 13-month smoothed Rz) was about 4 years. From the scatter diagram it can be noticed that the Cycles 3 (R$_{max}$= 157.8) and 22 (R$_{max}$ = 159.4) with lesser amplitudes compared to Cycle 19 have lower rise times (3 and 3.5 years respectively) indicating that the rise time of solar cycle will have certain minimum time of ascent and can not decrease monotonously with increase in R$_{max}$.    This effect is explained by the shift in the timing of minima: due to the overlap between consecutive cycles stronger,
 rapidly growing cycles reach their minimum early, thus increasing the time between minimum and maximum \cite{Cameron08}. This, in fact, exactly is seen in case of rise time of Cycle 19 wherein the reversal of the Waldmeier effect occurs.  However, neither anticorrelated linear fit (dotted and dashed lines) nor an inverse (T$_R$ Versus 1/R$_{max}$) linear fit (dash-dot-dot-dot curve in Figure 5) of the form T$_R$=2.15 + 206/R$_{max}$ (Hathaway, 2010) fits the reversal of Waldmeier effect. 

%
 \begin{figure} 
 \centerline{\includegraphics[width=0.6\textwidth,clip=]{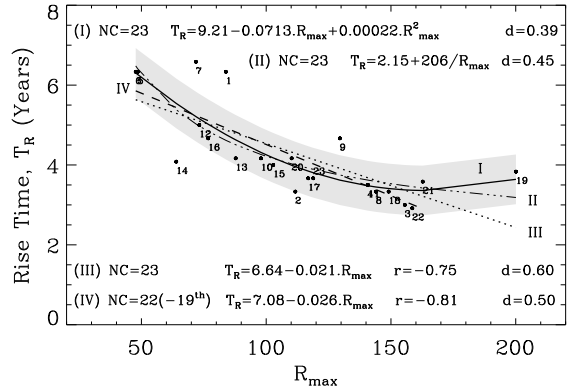}}
 \caption{T$_R$ Versus R$_{max}$.  Thick continuous line with label (I) and the dash-dot-dot-dot line with label (II) depict respectively the second order polynomial and the inverse fit of T$_R$ and R$_{max}$. Dotted line with label (III) and the dashed line with label (IV) depict respectively the linear fits of T$_R$ and R$_{max}$ of all the 23 cycles and 22 cycles excluding cycle 19. Corresponding fitted equations, number of cycles used (NC), correlation coefficients (if applicable) and mean absolute deviations are shown with respective labels.}
 \end{figure}

\inlinecite{Ramesh00} has suggested a second order polynomial fit between T$_R$ and R$_{max}$ that seems to have shown reasonably good prediction of time of maximum  (June-September 2000) for Cycle 23. A second order polynomial (thick continuous line) for all the 23 cycles is shown in Figure 5 and the equation of the  polynomial is given by T$_R$ (in years) = 9.21($\pm$1.04) - 0.0713($\pm$0.0190) R$_{max}$ + 0.00022($\pm$0.00008) R$_{max}^2$.  A simple test of goodness of fit using mean absolute deviation ([d=$\Sigma ( |$T$_i$-$\hat{T_i} | )$/nc] where T$_i$ is the cycle rise time, $\hat{T_i}$ is the fitted value of a particular statistical model and nc is the number of cycles) is performed.   The mean absolute deviation (d=0.39) of observations from the fitted curve for  the second order polynomial is less when compared to those of linear regressions (d=0.60 and 0.5 respectively for all the 23 cycles and for the line excluding Cycle 19) and the inverse relation (d=0.45). Therefore, the second order polynomial seems to be a better fit over the linear regression and the inverse relation.  Rise times of 19 out of 23 cycles are well within 1$\sigma$ level (Hatched region centered around the second order polynomial in Fig 5) of variation.  This curve, in fact, does not distinguish Cycle 19 as anomalous. It is important to note that there exists a minimum rise time of about 3 to 3.5 years that correspond to a  R$_{max}$ of about 160.  Further increase in R$_{max}$ leads to an increase in T$_R$ \cite{Cameron08} and the Waldmeier effect, decrease in T$_R$ with increase in R$_{max}$, breaks at this point.  In our opinion this possibility has not been pointed out in the earlier works related to the Waldmeier effect. This feature seems to persist even with T$_{RKC}$ that can be visualized from Fig 1 (Top left panel)  of \inlinecite{Karak11} wherein T$_{RKC}$ seem to saturate at about 1.4 years for higher R$_{max}$. The minimum rise time of 3-3.5 years probably help constraining the high turbulent diffusivity based flux transport dynamo model \cite{Karak11} which in turn may lead to more accurate model based predictions \cite{Choudhuri07}.  

\subsection{Prediction of  T$_R$ for cycle 24}

With the predicted R$_{max}$ of 85 the rise time for Cycle 24 is estimated to be 4.7 years.  From the smoothed Rz the time of R$_{min}$ is found to be during January 2009. Therefore, the time of occurrence of maximum of Cycle 24 will be during August 2013 and with the uncertainties of 1$\sigma$ level included the maximum of Cycle 24 may occur in the third quarter of the year 2013.  Prediction of time of R$_{max}$ for the upcoming solar cycle seems to be an approximate match  to the predictions of \inlinecite{Ahluwalia10} and \inlinecite{Kakad11}.

\section{Conclusions}

Strength of using R$_{max}$ and R$_{min}$ relationship for predicting the amplitude of the upcoming solar cycle lies in the fact that these quantities are directly linked through the internal dynamics of the Sun and that the errors in measuring them are of same origin.  Dependence of R$_{max}$ on R$_{min}$ and T$_R$ on R$_{max}$ are statistically viable and strengthening this method supports the view (Pesnell, 2008) that considering the solar and geomagnetic precursor methods separately would help assessing the overall predictions in a better way.  Minimum rise time of about 3-3.5 years with a moderate solar cycle strength of about 160 as measured in terms of sunspot number is an important finding of this paper. This effect may be due to the shift of the time of minima preceding stronger rapidly growing cycles which in turn increases the time between minimum and maximum. We believe, this may even constrain the flux transport dynamo models that would help revising them for more accurate physical principles based predictions.   Sunspot Cycle 24 will be of moderate strength with an R$_{max}$ of 85$\pm$17 and occur around  the third quarter of the year 2013.  Prediction of R$_{max}$ is similar to those of few others while the prediction of time of occurrence (third quarter of the year 2013) is a close match to that (June 2013) of Ahluwalia and Ygbuhay (2010) and \inlinecite{Kakad11}.

%

%

%

%
  \begin{acks}
Authors thank the reviewer for the constructive criticism and suggestions.
  \end{acks}

%
%
 \bibliographystyle{spr-mp-sola}
 \bibliography{kbr_paper17_ref.bib}  
%
%
%
%

\end{article} 
\end{document}